\documentclass{iopart}
\usepackage{graphicx}
\usepackage{amsfonts}
\usepackage{amssymb}
\begin{document}
\title{ Geometrical approach to SU(2) navigation with Fibonacci anyons}
\author{R\'emy Mosseri }
\address{Laboratoire de Physique Th\'eorique de la Mati\`ere Condens\'ee, CNRS UMR 7600,
Universit\'e UPMC Paris, 4 Place Jussieu, 75252 Paris Cedex 05, France }
\ead{mosseri@ccr.jussieu.fr; }

\begin{abstract}
Topological quantum computation with Fibonacci anyons relies on the possibility of efficiently generating unitary transformations upon pseudoparticles braiding. The crucial fact that such set of braids has a dense image in the unitary operations space is well known; in addition, the Solovay-Kitaev algorithm allows to approach a given unitary operation to any desired accuracy. In this paper, the latter task is fulfilled with an alternative method, in the $SU(2)$ case, based on a generalization of the  geodesic dome construction to higher dimension.   
\end{abstract}


\section{ Introduction}

Topological Quantum Computation (TQC) \cite{kitaev03, freedman02, preskill, dassarma07} makes use of the subtle properties of topological phases of matter to provide an original implementation for quantum computation, better immune to decoherence. Its main ingredients are anyonic excitations displaying non-abelian braiding statistics. Although no direct experimental proof exists that such characteristics occurs in real physical systems, there are some evidence that, for instance, the  $12/5$ Fractional Quantum Hall Effect states 
 should be good candidates to display the expected properties.

Up to now, contributions to the TQC field are mainly splitted into two parts, a "hardware" part whose main purpose is to find microscopic models, and possible experimental realizations displaying these topological features in their spectral properties, and a "software" part, which starts from a formal (non-abelian) anyon model, and defines, out of it, qubit states, quantum gates and algorithms. Notice that this splitting is already present in more "standard" qantum computation, with on one hand the large effort devoted to built experimental implementations of sets of coupled qubits, and the quantum algorithm part, which in fact started first, and most often do not discriminate between the very different microscopic realizations for the qubits, supposing that a large amount of them are already available.

In the present paper, we analyse a model with three Fibonacci anyons (irrespective of their implementation), and ask how their manipulation (upon braiding) can appropriately approximate the action of generic $SU(2)$ unitay transformations. As is well known (\cite{freedman02}), this is in principle possible to any desired accuracy, thank to the fact that the associated non-abelian braid group representation is dense in $SU(2)$. To make this system interesting, it is also important that this can be done efficiently. Such a task has yet been fulffilled \cite{bonesteel05, homozi07} by splitting the braid search into two distinct parts : first, a brute force search among all braids up to a given length to generate the closest matrix to the target one; then, a refinement step done by iteratively implementing the Solovay-Kitaev algorithm \cite{kitaev99,nielsen_chuang}. With additional Fibonacci anyons, it is possible to define more qubits, whose interaction results from appropriate braiding.  For example, a universal set of quantum gates has been derived \cite{bonesteel05,homozi07} , with six anyons forming a two-qubit system, proving that it can in principle allows for quantum computation.

Here an alternative approach is proposed, of rather different nature, in order to generate the $SU(2)$ elements. Instead of first insisting on the dense $SU(2)$ covering generated by the Fibonacci braid group generators, we start by analysing how good the latter can approximate the generators of binary polyhedral $SU(2)$ finite subgroups. It comes out that the subgroup of higher order, the binary icosahedral group $Y$ with 120 elements, can indeed be very efficiently approached. Recalling the isomorphism between $SU(2)$ and the 3-dimensional sphere $S^3$, this already allows a fine grained description of $SU(2)$. Indeed, to the group $Y$ corresponds the regular polytope $\{3,3,5\}$ \cite{coxeter73,sadocmosseri99}, whose full symmetry group $G$ (discrete subgroup of $O(4)$), has order $14400$. This already leads to an efficient way of generating $14400$ $SU(2)$ unitary transformations, related by symmetry. 

We further show how to iteratively gets finer and finer meshes in $SU(2)$ by generating the so-called "geodesic hyperdomes", the analogues with one dimension more, of the celebrated families of geodesic domes which provide fine discrete approximations of the usual sphere $S^2$.

In a final part, a more "disordered" version of the latter step is described, which already provides an efficient speedup for "brute-like" search.

\section{Binary icosahedral group generation with Fibonacci anyons}

Fibonacci anyons are quasiparticles displaying non-abelian statistics upon braiding. We will not recall here the whole derivation of their properties, which can be found elsewhere \cite{freedman02, preskill, dassarma07,bonesteel05,homozi07}, but only summarize what is used in the present context. What we need here is an expression for the two generators of the associated (non-abelian) 2-dimensional representation of the braid group $B_3$. A close inspection of the braiding and fusion rules, taking into account the need to satisfy the so-called pentagon and hexagon equations, allows to find a set of generators.

As shown in ref.\cite{freedman02}, a qubit (2-level) system can be associated with three Fibonacci anyons, with a third state (called "non computational"), which is not coupled to the first two upon anyons braiding. We shall therefore focus  to the $SU(2)$ unitary action (up to a global phase) onto the qubit space. 

\subsection{Braid group generators for Fibonacci anyons}
Generally speaking, a representation of the braid group $B_n$ has $n-1$ generators $\sigma_j$ satisfying the following two simple relations
\begin{eqnarray}
\label{eq:braiding}
\sigma_i\sigma_j &=\sigma_j\sigma_i, \ \left|i-j\right|\geq2 \\
\sigma_j\sigma_{j+1}\sigma_j &=\sigma_{j+1}\sigma_j\sigma_{j+1}, \ 1\leq j \leq n-2 \nonumber
\end{eqnarray}

which already limits the set of possible $\sigma_j$ matrices. Fibonacci fusion rules constrains further this set, which eventually leads to the following unique (up to a phase) solution  in the $B_3$ case :

\begin{eqnarray}
\label{eq:fibo_generators}
\sigma_1= \left(
\begin{array}{cc}
	\exp{-7i\pi/10} & 0 \\ 0 &\exp{7i\pi/10}
\end{array}
	\right), \\
	\sigma_2= \left(
\begin{array}{cc}
	-\tau\exp{-i\pi/10} & -i\sqrt{\tau} \\ -i\sqrt{\tau} & 	-\tau\exp{i\pi/10}
\end{array}	\right)
\end{eqnarray}	

with $\tau=\left(\sqrt{5}-1\right)/2$ the inverse golden mean. Note that $\sigma_1$ and $\sigma_2$ both satisfy 

\begin{equation}
\label{eq:order_ten}
\sigma_1^{10}=\sigma_2^{10}=-1
\end{equation}

Now, any braid is represented as a product of the $\sigma_j$ generators. It allows also for an unambigous graphical presentation, where $\sigma_j$ is displayed as a crossing between braid lines $j$ and $j+1$ (see for example 
Figure \ref{fig:braid_graph_st}). A word of caution should be given here concerning the braid ordering. Braid words are literally given and drawn here, as usual, with time flowing from left to right. Quantum qubit states however are represented as column vectors acted on the left by unitary matrices. Therefore, to build the unitary matrix corresponding to a braid word requires to reverse the order from the braid word to the associated matrix product.

\subsection{Binary polyhedral groups : geometry and generators}

Due to the $2:1$ homomorphism between $SU(2)$ and $SO(3)$, discrete $SU(2)$ subgroups have a counterpart as point groups in $R^3$. Let us focus here on the binary tetrahedral $T$ (order 24), octahedral $O$ (order 48) and icosahedral $Y$  (order 120) groups. When viewed as elements of $S^3$, $T$ and $Y$ correspond to the regular polytopes $\{3,4,3\}$ and $\{3,3,5\}$ \cite{coxeter73, sadocmosseri99}. The group presentations are given here together with sets of simple quaternions generators (see appendix A for a brief presentation of quaternions)

\begin{itemize}

\item {Binary tetrahedral group $T$}
\begin{eqnarray}
	<s,t|s^3&=t^3=(st)^2=-1> \\
	s&= (1+i+j+k)/2 \nonumber \\
	t&= (1+i+j-k)/2 \nonumber
\end{eqnarray}

\item {Binary octahedral group $O$}
\begin{eqnarray}
	<s,t|s^3&=t^4=(st)^2=-1> \\
	s&= (1+i+j+k)/2 \nonumber \\
	t&= (1+i)/\sqrt{2} \nonumber
\end{eqnarray}

\item {Binary icosahedral group $Y$}
\begin{eqnarray}
	<s,t|s^3&=t^5=(st)^2=-1> \\
	s&= (1+i+j+k)/2 \nonumber \\
	t&= (\tau^{-1}+\tau i+j)/2 \nonumber
\end{eqnarray}

\end{itemize}

\subsection{Binary icosahedral group approached with Fibonacci anyons}

We recalled above that the braid group $B_3$ representation with Fibonacci anyons is dense in $SU(2)$. We now question the possibility of generating the binary icosahedral subgroup by imposing some constraints on the words generated with  $\{\sigma_1,\sigma_2\}$. Trying to build new generators $\tilde{s}$ and $\tilde{t}$ from $\{\sigma_1,\sigma_2\}$, which would follow the above recalled generating relations for $Y$, we eventually find that, while two of the three relations are easily exactly fulfilled, the third one seems only asymptocally satisfied with longer and longer words. We call  these cases "`pseudo-generators" : a brute force search for best words up to length 10 already gives the following very good approximations :

A pseudo-generator $\tilde{s}=\sigma_2^2 \sigma_1^{-3}\sigma_2^2\sigma_1^{-1}\sigma_2\sigma_1$
\begin{eqnarray}
\tilde{s}=\left(
\begin{array}{ll}
 0.5-0.706298 i & -0.428519-0.2598349 i \\
 0.428519-0.2598349 i & 0.5+0.706298 i
\end{array}\right) \\
\mathrm{with} \ \tilde{s}^3=\left(
\begin{array}{cc}
 -1 & 0 \\
 0 & -1
\end{array}
\right) \nonumber
\end{eqnarray}
and a pseudo-generator $\tilde{t}= \sigma_1\sigma_2^2\sigma_1^{-2}\sigma_2\sigma_1^{-1}\sigma_2\sigma_1^{-1}\sigma_2$  

\begin {eqnarray}
\tilde{t}=\left(
\begin{array}{ll}
 -0.309017+0.159002 i & -0.414981+0.840843 i \\
 0.414981+0.840843 i & -0.309017-0.159002 i
\end{array}
\right)  \\
\mathrm{with} \ \tilde{t}^5=\left(
\begin{array}{cc}
 -1 & 0 \\
 0 & -1
\end{array}
\right) \nonumber
\end{eqnarray}

\begin{figure}[hc]
 \centering
\includegraphics [height=1in,width=2in]{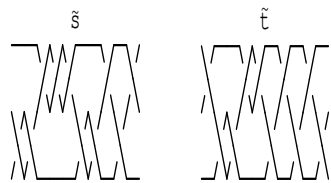}
\caption{\label{fig:braid_graph_st} Graphical representation for the pseudo-generators braids $\tilde{s}$ and $\tilde{t}$, built as words in the $\sigma_1$ and $\sigma_2$ Fibonacci generators. Note that $\sigma_1$ (resp. $\sigma_2$) refers to  crossing the upper (resp. the middle) braid with the middle (resp. the lower) braid, with the convention that the upper braid crosses "on top" of the lower braid (the reverse case coding the inverse $\sigma_1^{-1}$ and $\sigma_2^{-1}$). }
\end{figure}

These two pseudo-generators are shown on figure \ref{fig:braid_graph_st}. 
Note that in the above two expressions, the numerical values are cut up to 6 or 7 digits; but the $\tilde{s}$ and $\tilde{t}$ exact expressions, as products of the $\{\sigma_1,\sigma_2\}$ Fibonacci generators, are such that 
$\tilde{s}^3=\tilde{t}^5=-1$ is exact. Finally the third binary icosahedral group generating relation is only almost fulfilled
\begin{equation}
(\tilde{s}\tilde{t})^2=
\left(
\begin{array}{ll}
 -0.999995+0.000529 i & -0.001483-0.002677 i
   \\
 0.001483-0.002677 i & -0.999995-0.000529 i
\end{array}
\right)
\end{equation}

Now, it is easy to build, with short words in the pseudo-generators $\{\tilde{s},\tilde{t}\}$, a set denoted $\tilde{Y}$ corresponding to a very slightly deformed $\{3,3,5\}$ polytope. Since $\{\tilde{s},\tilde{t}\}$ do not exactly fulfill the $Y$ generating relations, their span is in principle infinite. What we are doing in fact is to select, once for all, $120$ words in $\{\tilde{s},\tilde{t}\}$ (e.g. $120$ braids) which very closely approximate the $Y$ elements. The word length never exceeds 8, which puts an upper bound of 80 to the length of the $\tilde{Y}$ elements in terms of the original $\{\sigma_1,\sigma_2\}$ Fibonacci generators. 

Note that it may still be possible to find shorter words leading to a good approximation of $Y$, either by the process of word contraction, or by finding equivalent approximations by brute force search in the original generators. We are not interested here in absolute length minimization, but rather to describe a fine grid mesh based on discrete subgroup and geodesic hyperdome iterative generation; we shall therefore stick to  $\{\tilde{s},\tilde{t}\}$ generated braids.

\section{ Iterative fine meshes in $SU(2)$ with geodesic hyperdomes}

We are going to build increasing sets $\mathcal{P}_i$ and $\mathcal{Q}_i$ which are the images under the full $G$ group of seed sets of points (denoted $\mathcal{S}_i^\mathcal{P}$ and $\mathcal{S}_i^\mathcal{Q}$) inside the orthoscheme $\mathcal{O}$ (the $G$ group fundamental region, see appendix C). The $\mathcal{P}_i$, having the form of "geodesic hyperdomes", were introduced, more than 20 years ago, in the very different context of atomic strutures with long range icosahedral order \cite{hierarc_poly84, hierarc_poly85}. More precisely, those $\mathcal{P}_i$ all shared the exact $G$ subgroup of $O(4)$, while here the sets $\mathcal{P}_i$ follows approximate symmetry operations  $\tilde{G}$, built from $\tilde{Y}$. But from now on, we shall no more differenciate the exact $Y$ and the approximate $\tilde{Y}$ in describing these sets. These polytopes $\mathcal{P}_i$ are built such that the vertices local order is very close to that of the $\{3,3,5\}$ vertices. In particular they have (slightly deformed) tetrahedral cells, each of which being decomposed into 24 smaller tetrahedra, which divide the larger tetrahedron in a way similar to the exact orthoscheme division of a $\{3,3,5\}$ cell (see appendix D, and for more details, ref. \cite{hierarc_poly84, hierarc_poly85}). The $\mathcal{Q}_i$ sets correspond to this finer division of the $\mathcal{P}_i$, with one generic point in each orthoscheme-like tetrahedra.

In order to generate, with Fibonacci anyons, the corresponding sets of unitary matrices, one proceeds as follows.
To get the full 14400 images (under $G$) of a generic matrix $q$ (noted as a unit quaternion) one must generate the elements (see appendix C) $l q r$ and $l \bar q r$ with $l,r\in Y$. In terms of braiding operations, $l$ and $r$ are, once for all, put in one-to-one correspondance with braids (also noted $l$ and $r$ for conveniance) written in the generators $\tilde{s}$ and $\tilde{t}$. The central braid associated with the (seed) matrix $q$ is then concatenated on the left and on the right by $l$ and $r$.

\subsection{The $\mathcal{P}_0$ and $\mathcal{Q}_0$ first meshes }

The first case is very simple, and directly associated with the binary icosahedral group $Y$. $\mathcal{P}_0$ corresponds to the $\{3,3,5\}$ polytope; the seed set $\mathcal{S}_0^\mathcal{P}$ is just an orthoscheme vertex, corresponding to one element of $Y$. $\mathcal{Q}_0$ is the maximal set invariant under the full $G$ group symmetry, and $\mathcal{S}_0^\mathcal{Q}$ contains one point inside the orthoscheme.

In order to represent the $SU(2)$ elements, we shall use a Hopf map from $SU(2)$ onto the complex plane, as explained in appendix B. Figure \ref{fig:hmappseudoY} (left) shows the Hopf maf of $\mathcal{P}_0$; the obtained orientation on $S^3$ is generic, which leads to a full Hopf map showing 60 distinct elements on the base space (a fibre containing only two opposite matrices $\pm M \in SU(2)$).  Only 51 among these 60 base points are shown here on a limited region. With the full Hopf map (with an inverse stereographic projection onto $S^2$), this set of 60 points forms a semi-regular polyhedron  with icosahedral symmetry. Figure \ref{fig:hmappseudoY} (right) displays the Hopf map of the  $\mathcal{Q}_0$ $14400$ elements. Note that this set, although much denser, still has some uncovered regions (of pseudo pentagonal shapes).

\begin{figure}[htc]
 \centering
\includegraphics [height=2in,width=4in]{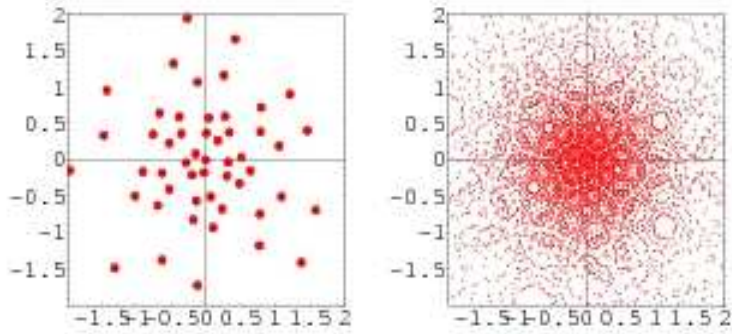}
\caption{\label{fig:hmappseudoY} Hopf map (onto the complex plane) of sets of $SU(2)$ elements. Left : the set $\tilde{Y}$ (e-g $\mathcal{P}_0$) with its 120 elements obtained from the pseudo-generators braids $\tilde{s}$ and $\tilde{t}$, which map onto 60 distinct points, one being sent to infinity and 51 being shown on this limited region. Right : Hopf map of the (14400 elements) set $\mathcal{Q}_0$.}
\end{figure}

\subsection {The finer meshes $\mathcal{P}_1$ and $\mathcal{Q}_1$ }
the set $\mathcal{P}_1$ corresponds to the first step of a geodesic hyperdome generation, as discussed in the appendix D. It contains 2160 points on $S^3$, which are the images under $G$ of $\mathcal{S}_1^\mathcal{P}$ made of three different points in the orthoscheme $\mathcal{O}$. 

One point coincides with a $\{3,3,5\}$ vertex (which is also an orthoscheme vertex), and can therefore be chosen conveniently as the identity matrix (in term of braiding operation, this means no braid in the seed region). The 119 other images can be simply taken by applying $Y$ either on the left or on the right. 

The second seed point in $\mathcal{O}$ is also an orthosheme vertex, located at the center of a $\{3,3,5\}$ tetrahedral cell; the whole 600 images under $G$ gives a $\{5,3,3\}$ polytope \cite{coxeter73}.

The third points sits along a $\{3,3,5\}$ edge, at 1/3 of the total edge length from a vertex. There are 1440 such points.

So, in order to generate $\mathcal{S}_1^\mathcal{P}$, one only need to generate two new $SU(2)$ matrices corresponding to these last two seed points. Approximating these two matrices with Fibonacci anyons is done by brute force search; reasonably good approximations are found upon inspection of all braids (in the $\{\sigma_1,\sigma_2\}$ initial Fibonacci generators) of limited length. Note that, since the full $G$ group is subsequently acted, it is not necessary that the initial brute force search generates the seed points  in the same $G$ group fundamental region; this point already improves greatly the speed of that search step, and will be a main ingredient of the alternative approach presented in paragraph 4.

The set $\mathcal{Q}_1$ is more complex to generate. $\mathcal{P}_1$ has 12000 (almost regular) tetrahedral cells. Each such cell can be subdivided into 24 smaller tetrahedra, in a way similar to the division of the perfect tetrahedral $\mathcal{P}_0$ cells into 24 orthoscheme copies : as a whole, $\mathcal{Q}_1$ has 288000 elements. The corresponding set $\mathcal{S}_1^\mathcal{Q}$ contains 20 points. Here again, the associated 20 $SU(2)$ matrices are generated by brute force search into finite length braids.

\subsection{The set of iterative finer meshes $\mathcal{P}_i$ and $\mathcal{Q}_i$}

The above construction of {$\mathcal{P}_1, \mathcal{Q}_1$} from {$\mathcal{P}_0, \mathcal{Q}_0$} can be be iterated { \it ad infinitum}. It can be seen simply as a site decoration procedure; it can also be derived from a barycentric construction detailed in ref \cite{hierarc_poly85}. We do not intend to recall this method here, and simply give in the table below some quantitative informations. Note that the number of sites grows by a constant factor ($20$) at each step in the sets $\mathcal{Q}_i$.

\begin{table}[ht]
\centering
\label{tab:hyperdomes}
\begin{tabular}{c | c | c | c | c | c | c | c }
$\mathcal{P}_0$  & $\mathcal{Q}_0$ & $\mathcal{P}_1$ & $\mathcal{Q}_1$& $\mathcal{P}_2$ & $\mathcal{Q}_2$ & $\mathcal{P}_3$ & $\mathcal{Q}_3$ \\ \hline
 $120$ & $14\,400$  & $2\,160$ & $288\,000$ & $42\,480$ & $5\,760\,000$ & $847\,440$ & $ 115\,200\,000$
\end{tabular}
\caption{ Number of sites ($SU(2)$ elements) in the first iterated hyperdomes $\mathcal{P}_i$ and related sets $\mathcal{Q}_i$}
\end{table}

\section {An alternative method to $SU(2)$ discretization}
\begin{figure}[htc]
 \centering
\includegraphics [height=2in,width=2in]{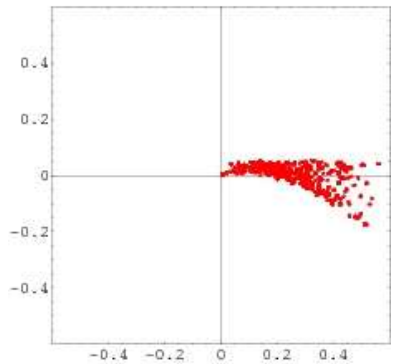}
\caption{\label{fig:O_fill} Hopf map (onto the complex plane) of $SU(2)$ matrices obtained from all words (up to length 7) in the  $\{\sigma_1,\sigma_2\}$ Fibonacci generators, and brought back, modulo the $G$ group action, in the same orthoscheme $\mathcal{O}$.}
\end{figure}

The above iterative hyperdomes correspond to almost regular coverings of $SU(2)$. One can also proceed differently, and get, rather efficiently, less regular coverings. We know, from the (once for all built) set $\tilde{Y}$, how to send any of the $14400$ $G$  fundamental regions onto a given one $\mathcal{O}$; we can  therefore focus on the filling operation limited to $\mathcal{O}$. This can be done by considering any matrix generated from a word in the braid generators $\{\sigma_1,\sigma_2\}$ . This $SU(2)$ element is in most cases outside $\mathcal{O}$; but it can be sent to $\mathcal{O}$ by the appropriate $G$ element. Generically, each new word therefore brings a new element in $\mathcal{O}$. Figure \ref{fig:O_fill} shows such ortoscheme filling for all words up to length 7. Applying the $G$ group 14400 elements (by concatenating braids on the left and on the right with the known 120 $Y$ elements) eventually leads to an already quite dense $SU(2)$ covering.

As an example, let us consider the the $SU(2)$ matrix $i\sigma_x$, which was approximated along a brute force search in ref. \cite{bonesteel05,homozi07}, where a braid of length 44 is found at a distance of about $10^{-3}$ of $i\sigma_x$. Here a  solution, equivalent under the $G$ symmetry group, and with the same order of magnitude accuracy, can be found with a braid of half this length, which then strongly reduces the brute force search. Note however that the full braid (with the $G$ elements acting on left and right) will eventually be longer than 44.

\section {Conclusions and comments}

Topological Quantum Computation with Fibonacci anyons strongly relies on the possibility of closely approaching any unitary matrix upon braiding the anyons. In this paper, we have shown how to fulfill this task for the "three anyons-one qubit" case, by generalizing the standard geodesic dome covering of the sphere $S^2$ to the "$SU(2)-S^3$" hyperdome case. The efficiency of this construction is due to the close, and yet unexplored, relation between Fibonacci braid generators and the binary icosahedral group generators. As a consequence, iterative finer and finer $SU(2)$ meshes can be generated, in a controlled way, with braid words of limited length. 

Generalization to many qubits (with more Fibonacci anyons) is not easy. The first step would consist in selecting the high order discrete subgroup of $SU(N)$ and try to approach their generating set by braiding the anyons. As usual, one should first focus on one and two-qubit gates, since it is known that generic $SU(N)$ can be generated by their suitable concatenation.  So the natural next step will be to analyse the "two-qubits $SU(4)$"  case.

One way, presently still under study, is to first analyse nice discrete sets  of two-qubits related by symmetry, and simply associated to successive shells of the eight-dimensional dense lattice $E8$ \cite{rigetti}. The first shell, with 240 points, corresponds (upon modding out a global phase factor) to 60 two-qubits "physical" states : 36 product states, and 24 maximally entangled (EPR) states. The product states are easily generated by separately braiding two sets of three braids (one needs only to use elements from the binary tetrahedral group, a $Y$ subgroup). The entangled states will require more subtle braiding operations, such that they keep the system inside the two-qubit Hilbert space.  Taking advantage of known properties about $E8$ shellings \cite{sadocmosseri93} (together with the entanglement sensitive $S^7$ Hopf fibration \cite{mosseridandoloff01}), larger sets of two-qubit states with intermediate entanglement could also be generated.

\bigskip

\ack{Acknowledgement}
It is a pleasure to acknowledge discussions with Pedro Ribeiro, Raphael Voituriez and David Bessis. This work (except the alternative method described in paragraph 4) was presented at the Dublin workshop on topological quantum computing in september 2007, where I benefitted from remarks by D. Bonesteels and M. Freedman.

\appendix
\section{ $SU(2)$ matrices, quaternions and the $S^3$ sphere}

Quaternions are usually presented with the imaginary units $\mathbf{i}%
,\mathbf{j}$ et $\mathbf{k}$ in the form~:
\begin{equation}
q=x_{0}+x_{1}\mathbf{i}+x_{2}\mathbf{j}+x_{3}\mathbf{k},\qquad x_{0}%
,x_{1},x_{2},x_{3}\in\mathbb{R}
\end{equation}

 with $ \mathbf{i}^{2}=\mathbf{j}^{2}=\mathbf{k}^{2}=\mathbf{ijk}=-1$,
the latter \textquotedblleft Hamilton\textquotedblright\ relations defining
the non-commutative quaternion multiplication rule. The conjugate of a quaternion $q$ is $\overline{q}=x_{0}-x_{1}\mathbf{i}%
-x_{2}\mathbf{j}-x_{3}\mathbf{k}$ and its
squared norm reads $N_{q}^{2}=q\overline{q}$. The set of normed (or unit) quaternions will be denoted $Q$.

Quaternions can also be defined equivalently, using the complex numbers $c_{1}=x_{0}+x_{1}
\mathbf{i}$ and $c_{2}=x_{2}+x_{3}\mathbf{i}$, in the form $q=c_{1}
+c_{2}\mathbf{j}$, or equivalently as an ordered pair of complex numbers
satisfying
\begin{eqnarray}
\left(  c_{1},c_{2}\right)  +\left(  d_{1},d_{2}\right)   &  =\left(
c_{1}+d_{1},c_{2}+d_{2}\right) \\
\left(  c_{1},c_{2}\right)  \left(  d_{1},d_{2}\right)   &  =\left(
c_{1}d_{1}-c_{2}\overline{d_{2}},c_{1}d_{2}+c_{2}\overline{d_{1}}\right)
\end{eqnarray}

Generic $SU(2)$ matrices read
\begin{equation}
M=\left(
\begin{array}{cc}
a+\mathbf{i}b & c+\mathbf{i}d \\
-c+\mathbf{i}d &a-\mathbf{i}b
\end{array}
\right), \ \mathrm{with} \ \ a^2+b^2+c^2+d^2=1 
\end{equation}

The latter relation (unit determinant) identifies $SU(2)$ to the the 3 dimensional sphere $S^3$.
Writing $M$ as

\begin{equation}
M=a\left(
\begin{array}{cc}
1 & 0 \\
0 & 1
\end{array}
\right)
+b\left(
\begin{array}{cc}
\mathbf{i} & 0 \\
0 &-\mathbf{i}
\end{array}
\right)
+c \left(
\begin{array}{cc}
0 & 1 \\
-1 & 0
\end{array}
\right)
+d \left(
\begin{array}{cc}
0 & \mathbf{i} \\
\mathbf{i} & 0
\end{array}
\right)
\end{equation}

allows to write $M$ as the unit quaternion 
\begin{equation}
M=a+b\mathbf{i}+c\mathbf{j}+d\mathbf{k},
\end{equation}
with the identification
\begin{equation}
\mathbf{i}\equiv\left(
\begin{array}{cc}
\mathbf{i} & 0 \\
0 &-\mathbf{i}
\end{array}
\right), \
\mathbf{j}\equiv \left(
\begin{array}{cc}
0 & 1 \\
-1 & 0
\end{array}
\right),\
\mathbf{k}\equiv \left(
\begin{array}{cc}
0 & \mathbf{i} \\
\mathbf{i} & 0
\end{array}
\right)
\end{equation}

\section{ Hopf map representation of $SU(2)$ matrices}

A fibred space $E$ is defined by a (many-to-one) map from $E$ to the so-called
\textquotedblleft base space\textquotedblright, all points of a given fibre
$F$ being mapped onto a single base point. A fibration is said "trivial" if
the base $B$ can be embedded in the fibred space $E$, the latter being
faithfully described as the direct product of the base and the fibre (think
for instance of fibrations of $R^{3}$ by parallel lines $R$ and base $R^{2}$
or by parallel planes $R^{2}$ and base $R$).

The simplest, and most famous, example of a non trivial fibration is the Hopf
fibration \cite{hopf} of $S^{3}$ by great circles $S^{1}$ and base space $S^{2}$.  One standard
notation for a fibred space is that of a map$E\stackrel{F}{\rightarrow}B$,
which reads here $S^{3}\stackrel{S^{1}}{\rightarrow}S^{2}$. Its non trivial
character implies $S^{3}\neq S^{2}\times S^{1}$.

To describe this fibration in an analytical form, we define elements 
of $S^{3}$ as pairs of complex numbers $\left(  \alpha,\beta\right)  $ which
satisfy $\left\vert \alpha\right\vert ^{2}+\left\vert \beta\right\vert ^{2}%
=1$. 

The Hopf map is defined as the composition of a map $h_{1}$ from $S^{3}$
to $R^{2}$ $(+\infty)$, followed by an inverse stereographic map $h_{2}$ from $R^{2}$ to $S^{2}:$%

\begin{eqnarray}
h_{1}  &  :%
\begin{array}
[c]{ccc}%
S^{3} & \longrightarrow & R^{2}+\left\{  \infty\right\} \\
\left(  \alpha,\beta\right)  & \longrightarrow & C=\alpha\beta^{-1}
\end{array}
\qquad\alpha,\beta\in\mathbb{C}\nonumber\\
h_{2}  &  :
\begin{array}
[c]{ccc}
R^{2}+\left\{  \infty\right\}  & \longrightarrow & S^{2}\\
C & \longrightarrow & M(X,Y,Z)
\end{array}
\qquad X^{2}+Y^{2}+Z^{2}=1
\end{eqnarray}

 The first map $h_{1}$ clearly shows that the full $S^{3}$ great circle, parametrized by ($\alpha\exp
i\omega,\beta\exp i\omega$), is mapped onto the same single point with complex
coordinate $C$. The Hopf map is therefore a mean to represent $SU(2)$ matrices, either on the complex plane or on the sphere $S^2$, but with identical images for matrices differing only upon multiplication by the matrix

\begin{equation}
\left(
\begin{array}{cc}
 \exp{\mathbf{i}\omega}& 0 \\
0 & \exp{-\mathbf{i}\omega}
\end{array}
\right)
\end{equation}

\section{ Polytope \{3,3,5\}}

Let us first recall the $\{p,q\}$ and $\{p,q,r\}$  Schl\"affli notations.
$\{p,q\}$ denotes a regular two-dimensional tiling (either spherical, euclidean or hyperbolic), such that each site belongs to $q$ regular $p$-gones : $\{4,3\}$ is a cube, $\{6,3\}$ is a honeycomb tiling.
$\{p,q,r\}$ is a regular three-dimensional  tiling, such that each edge belongs to $r$ polyhedra of the type $\{p,q\}$ : $\{4,3,4\}$ is the standard cubic tiling in $R^3$.

So, $\{ 3,3,5 \} $ denotes a tiling of regular tetrahedra $\{3,3\}$, with exactly five such tetrahedra sharing an edge. The regular tetrahedron dihedral angle being slightly less than $2\pi/5$, this leads to a polytope structure on the 3 dimensional curved space $S^3$ (embedded in $R^4$)\cite{coxeter73, sadocmosseri99}.
It contains :\hfill \break
\begin{itemize}
\item 120 vertices,  \hfill \break
\item 720 edges,  \hfill \break
\item 1200 triangular faces,  \hfill \break
\item 600 tetrahedral cells \hfill \break
\end{itemize}

Notice that these numbers satisfy the ($S^3$) generalized Euler-Poincar\'e relation, 
\begin{equation} 
V-E+F-C=0  
\end{equation}
where $ V,E,F,C$ are (respectively) the number of vertices, edges, faces and cells.

With one vertex on the pole, the successive "horizontal" sections are (i) an icosahedral shell,(ii) a dodecahedral shell, (iii) a new icosahedral shell, (iv) an equatorial icosidodecahedral shell. The next shells then symmetrically reproduce the same pattern down to the $S^3$ south pole.

The dual polytope $\{ 5,3,3\} $ has 600 vertices and 120 dodecahedral cells.

The $ \{ 3,3,5 \} $ symmetry group plays an important role in the present study. 
$S^3$ orientation preserving point symmetries form the group $SO(4)$, while the full group is $O(4)$. Symmetry elements are easily written in terms of unit quaternions. For the $SO(4)$ action, a given point on $S^3$, labelled by  the quaternion $q$, is sent to $lqr$, with $l,r \in Q $ (with an additional quotient by $Z_2$, see below). The remaining indirect symmetries in $O(4)$ are such that $q$ is sent to $ l \bar q r$.

The (properly oriented) $ \{ 3,3,5 \} $ 120 vertices (on a unit radius $S^3$) are in one-to-one correspondance with the 120 elements of the binary icosahedral group $Y$. Due to the group structure, multiplying on the left or on the right by $Y$ elements sends the polytope onto itself. Recalling that the group center is just $\{1,-1\}$, one finds as a whole the 7200 elements of the orientation preserving group $G'$ (discrete subgroup of $SO(4)$). 

\begin{equation}
G'=Y \times Y /Z_2 . 
\end{equation}

The full group $G$ includes 7200 additional indirect transformations, which reads
\begin{equation}
q \rightarrow l \bar q r\quad l,r\in Y ,
\end{equation}
leading as a whole to the $G$ 14400 elements.

This order can also be computed directly from the number of fundamental regions; for a regular polytope, this amounts to generate the tetrahedral orthoscheme $\mathcal{O}$ associated with the full symmetry group, such that the latter is generated by reflections about the orthoscheme faces. One orthoscheme is simply build from a regular cell $ \{ p,q \} $ of the $ \{ p,q,r \} $, by selecting a cell vertex $V$, a middle edge point $E$ (for an edge through the selected vertex), a middle face point $F$ (for a face sharing the cell vertex and the selected edge) and finally the cell centre C (see Figure \ref{fig:orthosch}-left)

 Polytope $ \{ 3,3,5 \} $ has 600 tetrahedral cells. Each cell being decomposed into 24 orthoschemes, one recovers
 the 14400 fundamental regions and therefore the $G$ group order. If one let the $G$ generators freely act onto a point $M$ in one orthoscheme $\mathcal{O}$, one eventually gets a set $\mathcal{P}$ of $N$ regularly spaced points on $S^3$, with $N$ depending on the location of $M$ :

 \begin{itemize}
\item If $M$ coincides with $V$, $N=120$ and  $\mathcal{P}$ is a  $ \{ 3,3,5 \} $ polytope \hfill \break
\item If $M$ coincides with $C$, $N=600$ and  $\mathcal{P}$ is a  $ \{ 5,3,3 \} $ polytope \hfill \break
\item If $M$ is  a generic point on a $\{ 3,3,5 \}$ edge, $N=1440$ , while $N=720$ if $M$ at a mid-edge position  \hfill \break
\item For a generic $M$ inside $\mathcal{O}$, the number of images is maximal, $N=14400$  \hfill \break
\end{itemize}

 Finally, as discussed in the text (and in the next appendix), one also considers sets $\mathcal{P}$ which are the image under $G$ of several points $M$, forming a seed set $\mathcal{S}$
 
\section{ Geodesic hyperdomes}

\begin{figure}[ht]
 \centering
\includegraphics [height=2cm]{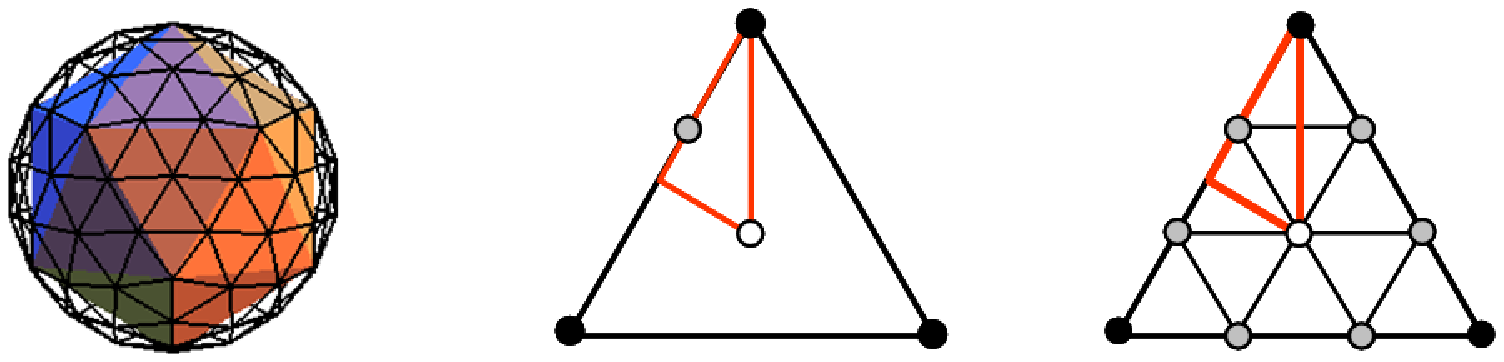}
\caption{\label{fig:dome2d} Left : geodesic dome based on icosahedral symmetry; center : an icosahedron triangular cell,  with a (fundamental region) orthoscheme decorated with three vertices (the so-called seed set $\mathcal{S}$) : a triangular cell vertex $V$ (black circle) , a face center $F$ (white circle), a vertex $D$ at one third on an edge(grey circle);  right : triangle cell decoration for  obtained as the local images of the three points in the orthoscheme }
\end{figure}

Geodesic domes are triangulations of the sphere $S^2$, usually built with icoshaedral symmetry. There are several different families of such discrete sets, the simplest being obtained by a decomposition of an icosahedron triangular faces into smaller triangles. Fig \ref{fig:dome2d}-left shows an example with 92 vertices, where edges are scaled by a factor 3 (this factor is only approximate if the dome vertices and edges are centrally mapped onto the sphere $S^2$). The geodesic dome shares the same symmetry group as the original icosahedron. Its vertices can therefore be generated from a seed set $\mathcal{S}$ located in one of the group orthoscheme. Figure \ref{fig:dome2d}-centre displays such an orthoscheme, inside a triangular face, with $\mathcal{S}$ made of three points, a face vertex $V$, face centre $F$ and a point $D$ at one third along an edge. The seed set is then propagated under the group action, here a reflection in the orthoscheme edges, leading to the geodesic dome 92 vertices in the following way : V has 12 images (forming the original icosahedron), F has 20 images (forming the dual dodecahedron), and $D$  has 60 images (forming a "buckyball" polyhedron). Figure \ref{fig:dome2d}-right shows the image of $\mathcal{S}$, propagated inside one triangle of the original icosahedron.

\begin{figure}[htc]
 \centering
\includegraphics [height=1.33in,width=4in]{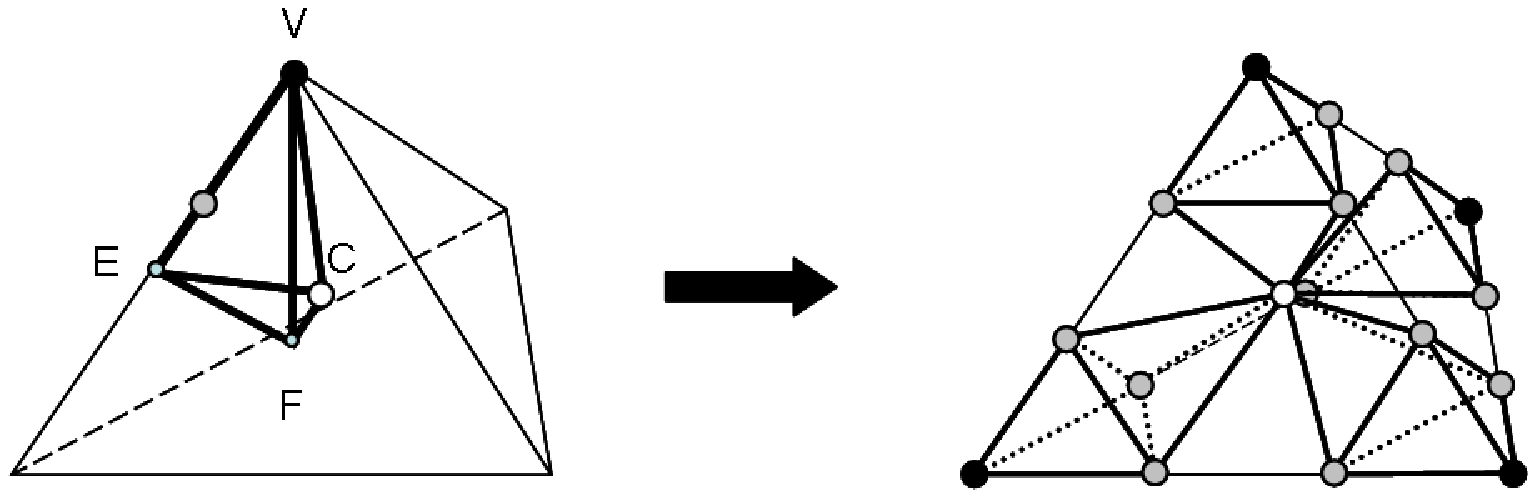}
\caption{\label{fig:orthosch} Left: a tetrahedral $ \{ 3,3,5 \} $ cell, with one fundamental orthoscheme whose four vertices are a cell vertex $V$, a mid-edge point $E$, a face center $F$, and the cell center $C$. The figure also shows the decoration of the orthoscheme by the seed set $\mathcal{S}_1^\mathcal{P}$, with $V$ (black circle), $C$ (white circle), and a third point located at one third on an edge (grey circle). Right : The cell decoration for $\mathcal{P}_1$, obtained as the local images of $\mathcal{S}_1^\mathcal{P}$ }
\end{figure}

The generalization to $S^3$ proceeds along similar lines. Take a $\{3,3,5\}$ tetrahedral cell (Figure \ref{fig:orthosch}-left), with one orhoscheme, and the three seed vertices described in paragraph 3-2. And then propagate the seed set under the $G$ group action. Figure \ref{fig:orthosch}-right shows the image of the propagated seed set, restricted to a $\{3,3,5\}$ tetrahedral cell.

\end{document}